\documentclass[12pt]{iopart}
\usepackage{listings}
\usepackage[draft]{minted}
\usepackage[caption=false]{subfig}
\usepackage{hyperref}
\usepackage{graphicx}
\usepackage[utf8]{inputenc}
\usepackage{tikz}
\usepackage[labelfont=bf]{caption}
\usetikzlibrary{calc}

\newminted{cpp}{gobble=2, numbersep=3pt, escapeinside=||, xleftmargin=5mm, framesep=2mm, frame=leftline, baselinestretch=1, fontsize=\scriptsize, linenos=true}
\newmintinline{cpp}{}

\makeatletter
\AtBeginEnvironment{minted}{\dontdofcolorbox}
\def\dontdofcolorbox{\renewcommand\fcolorbox[4][]{##4}}
\makeatother

\usetikzlibrary{shadows, arrows, fit, positioning, matrix, shapes}
\usetikzlibrary{decorations.markings}

\definecolor{my_yellow}{RGB}{255, 253, 217}
\definecolor{my_orange}{RGB}{255, 127, 0}
\definecolor{my_lightblue}{RGB}{105, 186, 249}
\definecolor{my_purple}{RGB}{150, 154, 219}
\definecolor{my_green}{RGB}{90, 194, 160}

\tikzset {
  bigbox/.style = {draw, thick, fill=gray!10, rounded corners, rectangle},
  box/.style = {draw, thick, minimum height=0.8cm, minimum width=1.5cm, rounded corners, rectangle, fill=white, anchor=south},
  model/.style = {draw, thick, fill=white, text centered, minimum height=3em, minimum width=4em, rounded corners, drop shadow},
  user/.style = {draw, thick, ellipse, fill=white, text centered, minimum height=3em, minimum width=5em, drop shadow},
  line/.style = {->, thick, color=black, shorten <=2pt, shorten >=2pt, >=stealth'},
  dashedline/.style = {->, thick, dashed, color=black, shorten <=2pt, shorten >=2pt, >=stealth'},
  plain/.style = {minimum width=1em},
  arcnode/.style 2 args={
    decoration={
                 raise=#1,             
                 markings,   
                 mark=at position 0.5 with {\node[inner sep=0] {#2};}
            },
            postaction={decorate}
    }
}

\pgfdeclarelayer{background}
\pgfdeclarelayer{foreground}
\pgfsetlayers{background,main,foreground}

\begin{document}

\title{Migrating large codebases to C++ Modules}

\author{Yuka Takahashi \textsuperscript{1}, Oksana Shadura \textsuperscript{2}, Vassil Vassilev \textsuperscript{3}}
\address{\textsuperscript{1} University of Tokyo, 7-3-1 Hongo, Bunkyo-ku, Tokyo 113-8654, Japan \textsuperscript{2} University of Nebraska-Lincoln, 1400 R St, Lincoln, NE 68588, USA \textsuperscript{3} Princeton University, Princeton, New Jersey 08544, USA}
\ead{yukatkh@is.s.u-tokyo.ac.jp}

\begin{abstract}
ROOT has several features which interact with libraries and require implicit header inclusion. This can be triggered by reading or writing data on disk, or user actions at the prompt. Often, the headers are immutable, and reparsing is redundant. C++ Modules are designed to minimize the reparsing of the same header content by providing an efficient on-disk representation of C++ Code. ROOT has released a C++ Modules-aware technology preview, which intends to become the default for the ROOT 6.20 release.

In this paper, we will summarize our experience with migrating C++ Modules to LHC experiment's software codebases, particularly for CMS software (CMSSW). We outline the challenges with adopting C++ Modules for CMSSW, including the integration of C++ Modules support in the CMS build system and we will evaluate the CMSSW performance benefits.
\end{abstract}

\section{Introduction}
\label{intro}

In High Energy Physics (HEP), experiments such as CMS, produce a large amount of data each second, and we expect to have even more data generated during the Hi-Lumi LHC \cite{hilumi} era. Thus software for HEP is always striving to archive better performance, which could improve the users data analysis as well the cost optimisation for the HEP resources used in the data centers.

ROOT~\cite{root} is a core software used in HEP not only for data analysis but also as a backend for LHC experiment's software such as CMSSW. Thus the performance improvement in ROOT will result in also improving performance of the experiment's software, which requires more resources than ROOT itself.

C++ Modules \cite{vassil-paper} has been supported in ROOT since version 6.16, in order to improve its performance. The cost of header re-parsing can be negligible in small to medium size codebases, but can be critical in larger codebases. In this case compile-time scalability will be affected, even though it will not affect the programs at runtime. However, ROOT is different -- its C++ interpreter Cling \cite{cling} processes code at program execution time and avoidance of redundant content re-parsing yields better runtime performance.

\section{Background}

\subsection{C++ Modules in ROOT's Dictionaries}
\label{intro}

C++ Modules represent the compiler internal state serialized on disk and deserialized on demand to avoid repetitions of parsing operations, as we have described in \cite{chep-modules}. There are different implementations of the concept in Clang, GCC, and MSVC. ROOT uses Clang through its interpreter Cling and adopts its C++ Modules implementation, which is one of the most sophisticated on the market as of today.

Many operations in ROOT require loading a shared library and its corresponding header files. For example, when we serialize or deserialize a ROOT file, ROOT needs to know which library contains how to stream its object. This meta information is widely known as ROOT "dictionary" information. The dictionary is generated by a platform-independent tool, \textit{rootcling}, which processes ROOT-aware code and produces a C++ source code file with a prefix \textit{G\_\_}. The file is later compiled and linked into the library it describes.

Generally, there are two ways to load a library: implicitly or explicitly. Both models require the library description to be available before or no later than library loading time. That is, all header files describing the library should be processed by the interpreter blindly without knowing if they will be used. This is a severe performance bottleneck which ROOT addresses with several technologies: \textit{ROOTMAP}s, \textit{RDICT}s, and a \textit{PCH}. The ROOTMAP file represents a lightweight version of the library descriptor containing forward declarations and their mapping to the corresponding library. The RDICT file represents a cache of a subset of \textit{TClass} objects which will be created when the library is loaded. The PCH file contains almost all of the headers in ROOT in an optimized form which is loaded at startup to avoid header re-parsing.

The ROOTMAPs and RDICTs are home-grown fixes to the inherent PCH problem -- it is impossible to extend a PCH without fully recompiling it. C++ Modules are the industry solution to this problem -- they represent composable PCH files called \textit{PCM}s.

A PCM is generated by rootcling using a special file containing the mapping between a header file and the module to be created, called a \textit{modulemap} file. It describes how a collection of existing headers corresponds to the logical structure of a module. It is important to note that transitive includes not present in a modulemap file are persisted multiple times. That is, if \textit{A.h} includes \textit{C.h}, \textit{B.h} includes \textit{C.h} and the modulemap maps A.h to module \textit{Alpha} and B.h to module \textit{Beta}, the content of C.h will be duplicated in both Modules. Such duplication can reduce performance (increasing parsing time) dramatically and should be avoided at almost any cost. This important detail suggests that "modularization" should be done bottom-up, namely, starting from external dependencies onward.

\subsection{C++ Modules in CMSSW}
\label{cmssw}

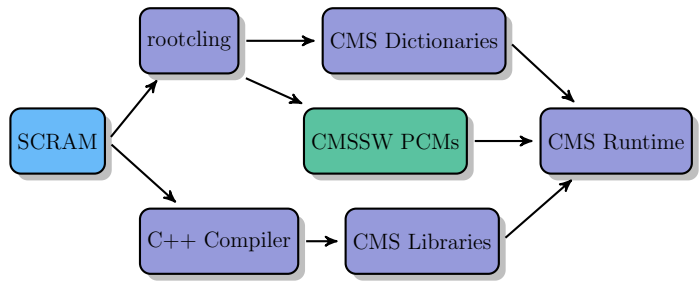
\begin{figure}[!h]
  \centering
  \begin{tikzpicture}[outer sep=0.05cm, node distance=0.8cm, scale=0.7, transform shape]
        
    \node[model, fill=my_lightblue, name=scram] (scram) {SCRAM};
    \node[model, fill=my_purple, name=compiler, below right=of scram] (compiler) {C++ Compiler};
    \node[model, fill=my_purple, name=rootcling, above right=of scram] (rootcling) {rootcling};
    \node[model, fill=my_purple, name=dicts, right=1.5cm of rootcling] (dicts) {CMS Dictionaries};
    \node[model, fill=my_purple, name=libs, right=of compiler] (libs) {CMS Libraries};
    \node[model, fill=my_purple, name=cms, above right=of libs, below right=of dicts] (cms) {CMS Runtime};
    \node[model, fill=my_green, name=pcm, left=1.3cm of cms] (pcm) {CMSSW PCMs};

    \draw[line, ->] (scram.east) -- (rootcling);
    \draw[line, ->] (scram.east) -- (compiler);
    \draw[line, ->] (compiler.east) -- (libs);
    \draw[line, ->] (rootcling.east) -- (dicts);
    \draw[line, ->] (dicts.east) -- (cms);
    \draw[line, ->] (libs.east) -- (cms);
    \draw[line, ->] (rootcling.south east) -- (pcm);
    \draw[line, ->] (pcm.east) -- (cms);
    
  \end{tikzpicture}
  \caption{Dependency Graph of C++ Modules in CMSSW}
  \label{fig:graphofcmssw}
\end{figure}

CMS \cite{cms} is one of the largest experiments in LHC and performance improvement of its software is crucial for the coming HL-LHC. CMS experiment develops its software stack called CMSSW (CMS SoftWare) which uses ROOT as its backend.

CMSSW utilizes their own build system, SCRAM \cite{scram}. SCRAM is a configuration management tool, a distribution system, a build system, and a resource manager, with local resources and applications managed transparently.

As shown in Figure \ref{fig:graphofcmssw}, the SCRAM build system resolves library dependencies and executes genreflex and C++ compiler such as gcc. Genreflex is a wrapper of rootcling, which supports legacy interface for experiments using ROOT. Rootcling generates CMS dictionaries such as rootmaps and RDICTs, as it was in ROOT. Simultaneously, C++ Compiler compiles CMS C++ files and generates CMS shared object libraries and CMSSW executable. Generated CMS libraries and dictionaries are being loaded at CMS Runtime.

We added C++ Modules to the existing CMS system, shown in Figure \ref{fig:graphofcmssw}. Rootcling generates CMSSW PCMs using a modulemap generated by SCRAM, and they will be used at CMS Runtime to save its parsing overhead.

\section{Migrating CMSSW to C++ Modules}
\label{migration}
Modularizing CMSSW requires two steps: generating a modulemap file and enabling C++ Modules in \textit{genreflex(rootcling)}. The generation of modulemap file happens at configuration time where we produce a file describing each library as a module and each library header file as a submodule. This can be subtle in some build systems as it requires tracking of header files which usually is not required. SCRAM already supports this and it was trivial to synthesize that description. Enabling C++ Modules in rootcling is done by adding an extra \textit{--cxxmodule} flag. We started doing that gradually library-by-library to avoid unnecessary complications.

The mixed mode support ensures an incremental migration path to C++ Modules while having a stable system throughout the migration period. For instance, the dictionaries of CMSSW may use the old dictionary system while the dictionaries of ROOT use the new technology. Ultimately, the mixed mode will not manifest into performance improvements because the C++ Modules technology intends to address dictionaries outside of the ROOT PCH, namely, in third-party software.

This section describes the steps taken in CMSSW in order to migrate to C++ Modules.

\subsection{Header Sanitizing}
In most cases, a module corresponds to a single dictionary or a library. Each module enumerates every header file in a submodule. Each submodule needs to be able to compile in isolation. Illustratively, a separate compiler instance is run on each header file. This assumes that every header file should be able to compile on its own. \textit{Standalone} header files include what they use and are resilient to configuration macros.

CMSSW codebase had a lot of header files inaccuracies thoroughly described in the GitHub C++ Modules Meta Issue~\cite{Modules-gh-metaissue}. The issues can be classified into the following categories:
\begin{itemize}
    \item Incomplete headers -- header files which do not include what they use. They are easy to fix because the C++ Modules system usually is able to suggest which are the omitted header files;
    \item Broken headers (unnecessary headers) -- header files which were never compiled. Those header files were never included in any translation unit but were "part" of a library. They are easy to deprecate and remove;
    \item Cyclic headers -- header files which include graph contains a cycle. For example, header \textit{A} includes header \textit{B} which includes \textit{A}. Header \textit{A} is in module \textit{Alpha} and header \textit{B} in \textit{Beta}. Usually this is a signal for a layering violation -- concepts from one library depend on another an vice versa. In many cases, using forward declarations of such entities resolves the problem. In some cases more sophisticated engineering techniques are necessary, such as, refactoring, moving the two dependent headers together or splitting the common dependent logic into its own library and module. In rare cases, if the mutual dependence of headers is by design they have to be in a single submodule.
    \item Macro headers -- header files which contain predominantly macro definitions which can be expanded differently in each translation unit. For example, \textit{$<$assert.h$>$} which conditionally defines the \textit{assert} if \textit{NDEBUG} is not defined. Macro headers are meant to be always textually included and they should be marked as \textit{textual headers} in the modulemap file. 
    \item Token generating headers -- header files which enable preprocessor metaprogramming should be excluded from the modulemap file.
\end{itemize}

After sanitizing the header files, a toolchain to protect regressions can be introduced. In CMSSW every header in a pull request is checked for the aforementioned issues. This is done by precompiling each header on its own. In future, we will deploy the include graph sanitization tool which we used to detect include cycles~\cite{scram-cycle-break}.

\subsection{Modularizing External Dependencies}

In order to fully see the expected performance benefits, we should modularize external dependencies as well. An external dependency should be modularized if a module transitively includes it. For example if header \textit{A} from module \textit{Alpha} includes directly or indirectly \textit{$<$vector$>$}, the external library, \textit{libstdc++}, containing that file should be modularized.

It can be challenging to modularize external dependencies. On one hand, it may be difficult to know what its modulemap file should contain, on the other hand, external dependencies can be located in non-writable locations on the file system. We have prepared a set of modulemap files for all external dependencies of CMSSW~\cite{raphael-auto-Modules}. The second challenge is solved by a file which instructs rootcling to pretend that the modulemap file is located at the non-writable folder in the file system. The \textit{virtual file system overlay file (VFSOF)} is programmatically synthesized by Cling.

\subsection{Automatic Generation of Modulemap Files and Virtual File System Overlay File}
\label{autogen}

As was mentioned in the section \ref{intro}, one of the important features to be adjusted is generation modulemaps. Well-behaved header files are trivial to enumerate in the modulemap file, as illustrated in Listing~\ref{list:modulemap}. The modulemap file syntax allows easy automation and most of its content can be generated by a build system. In CMSSW, modulemap autogeneration was simple due to its build system and header files structure. The codebase has the library interface headers files in a separate folder. We automatically generate the modulemap by iterating through those interface headers. It is expected that the modulemap is generated at configuration time but it must be generated no later then the first invocation of \textit{rootcling}.

\begin{listing}[h]
\noindent
\begin{minipage}[h]{\textwidth}
\begin{cppcode*}{linenos=false}
  module DataFormatsTrackerCommon_xr {
    module "TrackerTopology" {header "DataFormats/TrackerCommon/interface/TrackerTopology.h" export *}
    module "TrackerDetSide" {header "DataFormats/TrackerCommon/interface/TrackerDetSide.h" export *}
    module "ClusterSummary" {header "DataFormats/TrackerCommon/interface/ClusterSummary.h" export *}
  }
\end{cppcode*}
\end{minipage}
\caption{An example of a C++ Module definition in the CMSSW modulemap file.}
\label{list:modulemap}
\end{listing}
  
\begin{figure}[!h]
  \centering
  \begin{tikzpicture}[outer sep=0.05cm, node distance=0.8cm, scale=0.7, transform shape]
        
    \node[model, fill=my_lightblue, name=root] (root) {/};
    \node[model, fill=my_lightblue, name=usr, below right = 3cm of root] (usr) {usr/};
    \node[model, fill=my_lightblue, name=include, right = 2cm of usr] (include) {include/};
    \node[model, fill=my_lightblue, name=cpp, right = 2cm of include] (cpp) {c++/};
    \node[model, fill=my_lightblue, name=build, above right = 3cm of root] (build) {builddir/};
    \node[model, fill=my_lightblue, name=inc, right=2cm of build] (inc) {include/};
    \node[model, fill=my_lightblue, name=somedir, right=2cm of inc] (somedir) {somedir/};
    \node[model, fill=my_lightblue, name=headers, right = 1cm of somedir] (headers) {*.h};
    \node[model, fill=my_purple, name=stlmodmap, below left = 2cm of inc] (stlmodmap) {stl.modulemap};
    \node[model, fill=my_purple, name=libcmodmap, right = 1cm of stlmodmap] (libcmodmap) {libc.modulemap};    
    \node[model, fill=my_purple, name=modmap, right=1cm of libcmodmap] (modmap) {module.modulemap};
    \node[model, fill=my_purple, name=virtmap, above = 1cm of include] (virtmap) {VFSOF};

    \draw[line, ->] (root.east) -- (usr);
    \draw[line, ->] (usr.east) -- (include);
    \draw[line, ->] (include.east) -- (cpp);
    \draw[line, ->] (root.east) -- (build);
    \draw[line, ->] (build.east) -- (inc);
    \draw[line, ->] (inc.east) -- (somedir);
    \draw[line, ->] (somedir.east) -- (headers);
    \draw[line, ->] (inc.south) -- (stlmodmap);
    \draw[line, ->] (inc.south) -- (modmap);
    \draw[line, ->] (inc.south) -- (libcmodmap);
    \draw[dashedline, ->] (libcmodmap.south) -- (virtmap);
    \draw[dashedline, ->] (stlmodmap.south) -- (virtmap);
    \draw[dashedline, ->] (virtmap.south) -- (include);
    \draw[dashedline, ->] (virtmap.south) -- (cpp);

  \end{tikzpicture}
  \caption{Modulemap directory physical vs virtual structure.}
  \label{fig:modulemap}
\end{figure}
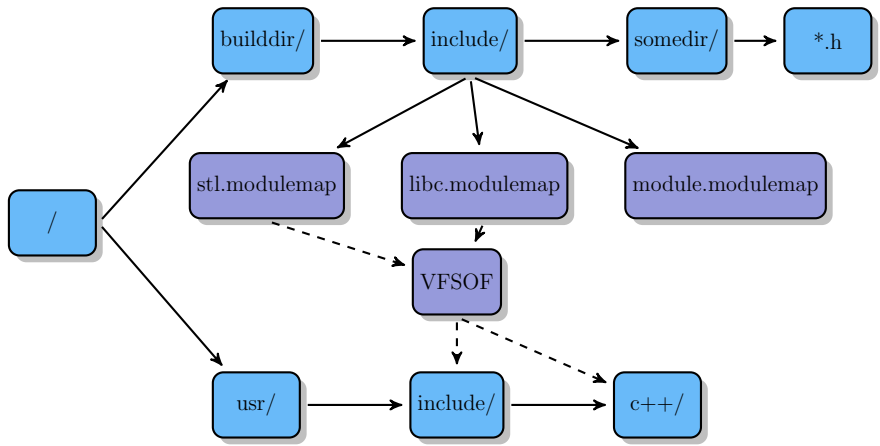

The process of modularization subsumes external dependencies such as \textit{libc}, \textit{libstdc++} and \textit{libxml}. Usually, external libraries do not ship with modulemap files. Listing~\ref{list:modulemap} demonstrates that the header location is relative to the include path in modulemaps. However, system headers are located in non-writable areas and modulemaps are impossible to be placed in order to access the headers with relative paths, unless super user privileges are granted.

\begin{listing}[h]
\noindent
\begin{minipage}[h]{\textwidth}
\begin{cppcode*}{linenos=false}
  { 'version': 0,
    'roots': [
      { 'name': '/usr/include/c++/', 'type': 'directory',
        'contents': [
          { 'name': 'module.modulemap', 'type': 'file',
            'external-contents': '/builddir/include/stl.modulemap' }]},
      { 'name': '/usr/include/', 'type': 'directory',
        'contents': [
          { 'name': 'module.modulemap', 'type': 'file',
            'external-contents': '/builddir/include/libc.modulemap'
        }]}]}
\end{cppcode*}
\end{minipage}
\caption{An example of a VFSOF for libc.modulemap and stl.modulemap.}
\label{list:vfsof}
\end{listing}

Figure \ref{fig:modulemap} shows the organization of the C++ Modules infrastructure together with the module map system. Blue rectangles describe the physical directories, and purple rectangles describes the physical files. Black thick arrows denote physical files and dashed lines show the virtual file system relationship.

Modulemap files for libc and stl (libstdc++) need to be present in \textit{/usr/include} and \textit{/usr/include/c++} respectively. Listing~\ref{list:vfsof} illustrates how the VFSOF instructs the infrastructure to consider \textit{libc.modulemap} and \textit{stl.modulemap} as if they were physically present in the expected folders. This approach allows us to "mount" modulemap files anywhere on the system enabling full software stack modularization.

The pre-configured VFSOF did not match the deployment process of CMSSW due to its static nature. The PCMs became non-relocatable from build directory as the path was hardcoded to PCMs. A more dynamic, on-memory representation of VFSOF was introduced to solve the problem. As it is a flexible virtual file on memory, paths do not need to be hardcoded at the configuration time, and thus it can determine header paths at runtime after the binary was distributed.

\section{Preliminary Performance Results}
\label{results}

The mixed run mode in ROOT allows us to make partial performance studies to assess the impact of this technology. The performance study was conducted on the CMSSW build server, where all the CMS software is continuously integrated. As CMSSW is large and there exist huge variations of possible physics workflows, we measured  realistic CMS workflow tests which were expected to represent the actual code that physicists would run for their analysis.

\begin{figure}[h]
\centering
\begin{minipage}{.48\textwidth}
\subfloat[] {\label{fig:perf:a} \includegraphics[width=\textwidth]{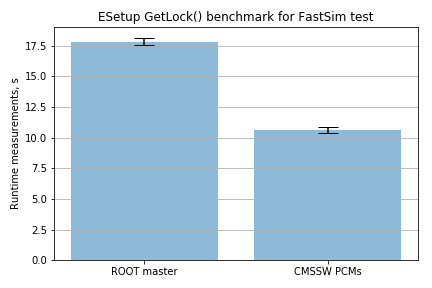}}
\end{minipage}\hfill
\begin{minipage}{.48\textwidth}
\subfloat[] {\label{fig:perf:b} \includegraphics[width=\textwidth]{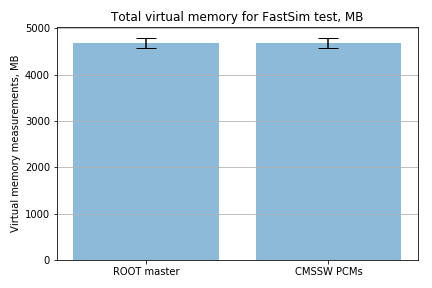}}
\end{minipage}
\begin{minipage}{.48\textwidth}
\subfloat[] {\label{fig:perf:c} \includegraphics[width=\textwidth]{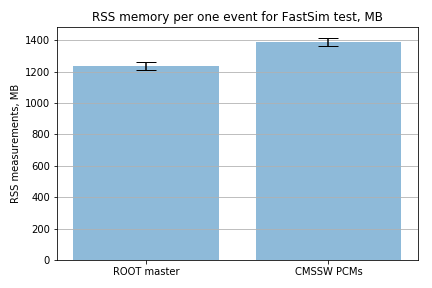}}
\end{minipage}\hfill
\begin{minipage}{.48\textwidth}
\subfloat[] {\label{fig:perf:d} \includegraphics[width=\textwidth]{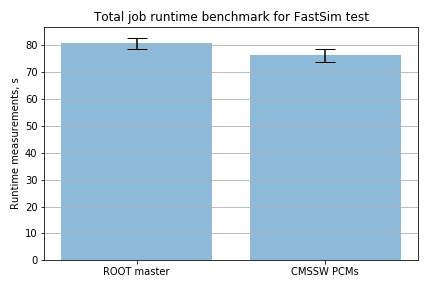}}
 \end{minipage}
 \caption{Performance results: (a) shows EventSetup GetLock() run time (RT) measurements for the fast simulation test. (b) shows  virtual memory measurements for the fast simulation test. (c) shows RSS measurements for the fast simulation test. (d) shows the total job measurements for the the fast simulation test.}
\label{fig:performance1}
\end{figure}

Figure \ref{fig:performance1} shows the benchmark of CPU time and RSS memory usage. On every plot, \textit{ROOT Master} is the case where ROOT and CMSSW built without C++ Modules, and serves as a baseline. \textit{CMS PCMs} is the configuration where ROOT (96 pcms) and CMSSW (25 pcms), which are both built with C++ Modules. Although CMSSW has over 200 libraries, we have modularized only 25 out of them at this point of the study.

 Figures \ref{fig:perf:c} and \ref{fig:perf:d} show the improvements for run time measurements, particularly for the event preparation phase (ESetup GetLock()). Figure \ref{fig:perf:d} proves the absence of performance degradation, which was the intent of the study at this stage of CMSSW modularization.  Figures \ref{fig:perf:b}, \ref{fig:perf:c} with memory measurements don't show the expected improvements. We miss memory performance improvements mainly due to two reasons: the incomplete modularization (only 15 \% of CMSSW libraries were modularized) and the preloading of all modules at startup time in sparse workflows, which is a reason why CMS PCMs use more RSS memory (see Figure \ref{fig:perf:d}). Error bars are caused by using different set of the events for CMSSW tests and possible fluctuations due to inconsistent CPU load of computing nodes used for tests.

\section{Conclusion}
We have shown the current advancements in the modularization of the CMSSW codebase. Despite the partial migration, we already observed some improvements. We are working towards modularization of the rest of the CMSSW components and external dependencies. We have implemented tools to aid this process and we are working on a further RSS memory reduction by avoiding preloading of all modules at startup time. It requires global module content indexing and loading the corresponding modules on demand. We expect more sophisticated benchmarking to be done, improving the coverage of many more workflows.

\section{Acknowledgments}
\label{ack}

This work has been supported by an Intel Parallel Computing Center grant, by U.S. National Science Foundation grants PHY-1450377, ACI-1450323 and PHY-1624356, and by the U.S. Department of Energy, Office of Science.

The authors are thankful to Shahzad Malik Muzaffar and Mircho Rodozov from the CMSSW development team, CERN/EP-SFT and the ROOT team.

\end{document}